\documentclass{article}
\usepackage{spconf,amsmath,graphicx}
\usepackage{amsmath,amssymb}
\usepackage{multirow}
\usepackage{flushend}
  \usepackage{textcomp}
  \usepackage[absolute,showboxes]{textpos}

  \TPMargin{5pt}
\newcommand{\copyrightstatement}{
    \begin{textblock}{0.84}(0.08,0.01)    
      \noindent
      \footnotesize
      \textcopyright~2022~IEEE.  Personal  use  of  this  material  is  permitted.
      Permission from IEEE must be obtained for all other uses, in any current or
      future media, including reprinting/republishing this material for advertising
      or promotional purposes, creating new collective works, for resale or
      redistribution to servers or lists, or reuse of any copyrighted component of
      this work in other works.
    \end{textblock}
  }

\title{non-autoregressive ASR with self-conditioned folded encoders}
\name{Tatsuya Komatsu}
\address{LINE Corporation, Japan}
\begin{document}
\copyrightstatement

%
\maketitle
\begin{abstract}
This paper proposes CTC-based non-autoregressive ASR with self-conditioned folded encoders.
The proposed method realizes non-autoregressive ASR with fewer parameters by folding the conventional stack of encoders into only two blocks; base encoders and folded encoders.
The base encoders convert the input audio features into a neural representation suitable for recognition.
This is followed by the folded encoders applied repeatedly for further refinement.
Applying the CTC loss to the outputs of all encoders enforces the consistency of the input-output relationship.
Thus, folded encoders learn to perform the same operations as an encoder with deeper distinct layers.
In experiments, we investigate how to set the number of layers and the number of iterations for the base and folded encoders.
The results show that the proposed method achieves a performance comparable to that of the conventional method using only 38\% as many parameters.
Furthermore, it outperforms the conventional method when increasing the number of iterations. 
\end{abstract}
\section{Introduction}
\label{sec:intro}
In recent years, end-to-end (E2E) automatic speech recognition (ASR) systems have achieved remarkable results~\cite{bahdanau2016end, chorowski2015attention, karita2019improving} owing to the drastic development of deep neural networks (DNNs).
Compared with the conventional hybrid systems~\cite{hinton2012deep}, which consist of acoustic, lexicon, and language models, E2E ASR simplifies its model architecture and procedures of both training and inference.
The most widely used model for E2E ASR is the attention-based encoder-decoder (AED) model~\cite{chorowski2015attention, dong2018speech}.
The AED model adopts an encoder-decoder architecture and generates tokens in an autoregressive manner: tokens are generated one by one from left to right.
Most of the state-of-the-art ASR systems~\cite{zhou2020self, gulati2020conformer} employ the AED model as a base architecture and have achieved promising results.
However, the AED model suffers from latency at inference time due to its autoregressive nature.
Another widely used E2E model is the connectionist temporal classification (CTC)~\cite{graves2006connectionist} model~\cite{kriman2020quartznet, baevski2020wav2vec}.
The CTC model consists of only stacked encoder layers and is trained to learn an alignment between input speech features and the output tokens.
CTC-based models have simple architecture and fast inference by parallel greedy decoding, but recognition accuracy is generally inferior to the comparable AED model due to the conditional independence assumption between output tokens.

A key factor in improving the performance of CTC-based models is how to handle dependencies between tokens in a non-autoregressive manner.
Approaches based on iterative refinement of the output tokens by token decoders are well known, and several methods have been proposed~\cite{chen2019listen, higuchi2020mask, chi2020align,higuchi2020improved}.
Mask-CTC~\cite{higuchi2020mask} is a model that inputs the output of the CTC to a decoder and refines the tokens of low confidence conditioned on the tokens of high confidence, and Align-Refine~\cite{chi2020align} is a model that inputs the potential alignment into the decoder and refines the model on the alignment space.
These models can be iteratively refined using special decoders to obtain dependency-aware output tokens. However, these methods require more complex model structures and training methods compared to standard CTC models. In addition, the incorporation of iterative decoding increases the computational complexity.

In contrast to iterative refinement methods, intermediate CTC~\cite{lee2021intermediate} and self-conditioned CTC~\cite{nozaki2021relaxing} have shown excellent performance using only CTC and audio encoders.
Intermediate CTC trains network parameters using the basic CTC loss and additional CTC losses for prediction at the intermediate layer.
Self-conditioned CTC is a simple extension of intermediate CTC that feeds back the predictions of the intermediate layer to the subsequent encoder.
A recent comparative study~\cite{higuchi2021comparative} reported that these methods showed the best performance among the latest non-autoregressive models, without any special decoder for the output tokens.

One of the reasons for the excellent performance of intermediate CTC and self-conditioned CTC is that they can perform the same operations as iterative token refinement by multiple stacked encoder layers.
Since intermediate CTC injects token information into the encoded representation by learning with intermediate CTC losses, we can say that the audio encoder is implicitly performing a transformation that includes dependencies between tokens.
Self-conditioned CTC makes explicit use of token information by the feedback structure.
Thus, these two models have no special decoders, but iterative refinements are performed implicitly/explicitly by stacked multiple encoder layers.
Looking over the two methods from this viewpoint, just as the token decoder can be used repeatedly, the audio encoder part may be built by the iterative use of a few encoder layers with shared parameters.

This paper proposes CTC-based non-autoregressive ASR with self-conditioned folded encoders.
The proposed method realizes non-autoregressive ASR with fewer parameters by folding the conventional stack of encoders.
By folding redundant encoders into encoders with shared parameters and applying them repeatedly, the proposed method performs the same operations as the conventional stack of encoders with fewer parameters.

\section{Overview CTC-based ASR method}
\subsection{Encoder for CTC-based ASR}
Let us consider a task that estimates a text sequence $\mathbf{y} \in \mathbb{\mathcal{V}}^{L}$ from an audio sequence $\mathbf{X}\in\mathbb{R}^{T\times{D}}$ , using $N$ stacked encoder layers and CTC~\cite{graves2006connectionist}.
Here, $\mathcal{V}$ and $L$ denote the vocabulary and length of the text sequence,
$D$ and $T$ denote the dimension and length of the audio sequence.
The shape of the input and output of each encoder is the same, and the output of one encoder is fed into the subsequent encoder as
\begin{equation}
    \mathbf{X}_{n} = \mathrm{Encoder}^{(n)} (\mathbf{X}_{n-1}),
\end{equation}
where $\mathbf{X}_{n-1}$ and $\mathbf{X}_n$ are the input and output sequences of the encoder, respectively.
The output of the final layer, $\mathbf{X}_N$, is mapped to the vocabulary domain by a linear layer, 
and then converted into the posterior distribution over the vocabularies $\mathcal{V}$ by applying the softmax function:
\begin{equation}
    \mathbf{Z}_{N} = \mathrm{Softmax}(\mathrm{Linear}_{D \rightarrow \mathcal{|V^{\prime}|}}(\mathbf{X}_{N})), \label{linear}
\end{equation}
where $\mathrm{Softmax}(\cdot)$ and $\mathrm{Linear}_{D \rightarrow \mathcal{|V^{\prime}|}}(\cdot)$ denote the softmax function and a linear layer that maps $D$ dimensional vectors to the $\mathcal{|V^{\prime}|}$ dimensional vectors.
$\mathbf{Z}_{N} \in \mathbb{R}^{T \times \mathcal{|V^{\prime}|}}$ is a posterior distribution over the vocabulary and $\mathcal{V}^{\prime} = \mathcal{V} \cup \{\epsilon\}$, where $\epsilon$ is a special blank token for CTC loss calculation.

For training, the CTC loss is calculated using $\mathbf{Z}_{N}$:
\begin{equation}
    \mathcal{L}^{\mathrm{CTC}} = -\log P_{\mathrm{CTC}}(\mathbf{y} \mid \mathbf{Z}_{N}).
\end{equation}
The CTC loss is defined as the log-likelihood over all valid alignments $\mathbf{a} \in \mathbb{\mathcal{V}^{\prime}}^{T}$ between $\mathbf{X}$ and  $\mathbf{y}$.
The log-likelihood is written as follows:
\begin{equation}
   \log P_{\mathrm{CTC}}(\mathbf{y} \mid \mathbf{X})=\log \sum_{\mathbf{a} \in \Gamma^{-1}(\mathbf{y})} P(\mathbf{a} \mid \mathbf{X})
\end{equation}
where $\Gamma$ is the function of $\mathbf{a}$, which removes all repeated tokens and special blank tokens.
The probability of each alignment $\mathbf{a}$ is modeled with the conditional independence assumption between tokens:
\begin{equation}
    P(\mathbf{a} \mid \mathbf{X}) = \prod_{t} P(a_{t} \mid \mathbf{X})
\end{equation}
where $a_t$ denotes the $t$-th symbol of $\mathbf{a}$.

\subsection{Intermediate CTC and Self-conditioned CTC}
Intermediate CTC~\cite{lee2021intermediate} is a method using additional constraints for CTC-based ASR, which can effectively improve the recognition accuracy.
In addition to the conventional CTC loss, the method calculates an additional CTC loss for the output of the intermediate layer.
The output of intermediate layer $\mathbf{X}_n$ is mapped to the vocabulary domain with the same linear layer in Eq.~\ref{linear} and intermediate prediction $\mathbf{Z}_n$ is obtained as follows: 
\begin{equation}
    \mathbf{Z}_{n} = \mathrm{Softmax}(\mathrm{Linear}_{D \rightarrow \mathcal{|V^{\prime}|}}(\mathbf{X}_{n})).
\end{equation}
Using this $\mathbf{Z}_n$, an additional constraint, intermediate loss, is calculated as
\begin{equation}
    \mathcal{L}^{\mathrm{Inter}} = - \frac{1}{|\mathbb{N}^\prime|}\sum_{n\in{\mathbb{N}^\prime}}\log P_{\mathrm{CTC}}(\mathbf{y} \mid \mathbf{Z}_{n}),
\end{equation}
where $\mathbb{N}^\prime$ is the set of layer indices for intermediate loss calculation.
The total loss of interCTC is defined as a weighted sum of the conventional CTC loss and the interCTC loss:
\begin{equation}
    \mathcal{L} = (1 - w) \mathcal{L}^{\mathrm{CTC}} + w \mathcal{L}^{\mathrm{inter}}, 
\end{equation}
where $w$ is a weight parameter.

Self-conditioned CTC~\cite{nozaki2021relaxing} is a simple extension of inter CTC by exploiting the intermediate representations $\mathbf{Z}_{n}$ for conditioning the subsequent encoder layers. 
During both training and inference, each intermediate prediction $\mathbf{Z}_{n}$ is fed back to the input of the next layer, 
\begin{align}
    \mathbf{X}_{n}^\prime &= \mathbf{X}_{n} + \mathrm{Linear}_{\mathcal{|V^{\prime}|} \rightarrow D}(\mathbf{Z}_{n})\\
    \mathbf{X}_{n+1} &= \mathrm{Encoder} (\mathbf{X}_{n}^\prime),
\end{align}
making the subsequent encoder layers conditioned on the intermediate predictions.
The loss for parameter training is same as interCTC.

In these methods, the mapping from audio representation to vocabulary domain in each layer, including the final layer, is performed by a single linear layer with shared parameters. 
In other words, the intermediate representations are transformed to follow distributions that can be discriminated by the same criteria based on the linear layer. 
In their structure, where the intermediate representation of one encoder layer is used for the input of the next layer, each encoder layer is considered to transform and refine the representation in the same space.

\section{The proposed method}
\subsection{Architecture}
The proposed method divides the encoder layers into two blocks: the base encoders and the folded encoders.
The base encoders and the folded encoders consist of $N_b$ and $N_f$ encoder layers, respectively.
First, the base encoders transform the input audio representation as in the conventional encoder layers,
\begin{align}
    \mathbf{X}_0 = \mathrm{BaseEncoders}(\mathbf{X}).
\end{align}
The folded encoders take the output of the base encoders and refine the audio representation,
\begin{align}
    \mathbf{X}_1 = \mathrm{FoldedEncoders}(\mathbf{X}_0).
\end{align}
This output is mapped to the posterior distribution of vocabularies, as in the conventional method, 
\begin{equation}
    \mathbf{Z}_{1} = \mathrm{Softmax}(\mathrm{Linear}_{D \rightarrow \mathcal{|V^{\prime}|}}(\mathbf{X}_{1})).
\end{equation}
Here, the proposed method feeds $\mathbf{Z}_1$ back into the audio domain as in self-conditioned CTC, 
and the folded encoders are applied repeatedly for further refinement
\begin{align}
    \mathbf{X}_{1}^\prime &= \mathbf{X}_{1} + \mathrm{Linear}_{\mathcal{|V^{\prime}|} \rightarrow D}(\mathbf{Z}_{1})\\
    \mathbf{X}_{2} &= \mathrm{FoldedEncoders} (\mathbf{X}_{1}^\prime).
\end{align}
By repeating this refinement $n_\mathrm{repeat}$ times, 
the proposed method perform the same operations as conventional encoders with  deeper  distinct  layers. 
After $n_\mathrm{repeat}$ iteration, the output $\mathbf{Z}_{n_\mathrm{repeat}}$ is used for the final prediction.

The reason for dividing the encoder into the base encoders and the folded encoders is that the input of the base encoder is generally raw, i.e., raw audio features or representation subsampled by CNNs.
It is not suitable for iterative transformation by the folded encoders with shared parameters.
Therefore, we first transform the first input by the base encoder without iteration and then refine the representation to achieve a stable iterative encoding.
Experiments evaluated the effect of the number of the base encoder.

The loss of the proposed method is the summation of the CTC losses for all the repeated outputs, 
\begin{equation}
    \mathcal{L}_\mathrm{repeat} = - \frac{1}{n_\mathrm{repeat}}\sum_{n = 1}^{n_\mathrm{repeat}}\log P_{\mathrm{CTC}}(\mathbf{y} \mid \mathbf{Z}_{n}).
\end{equation}
Here, $n_\mathrm{repeat}$ is an important parameter, and it can be a different value during decoding. 
Experiments also evaluated the impact of the number of iterations for both training and decoding.

\begin{figure}[tb]
\centering
\includegraphics[width=0.65\linewidth]{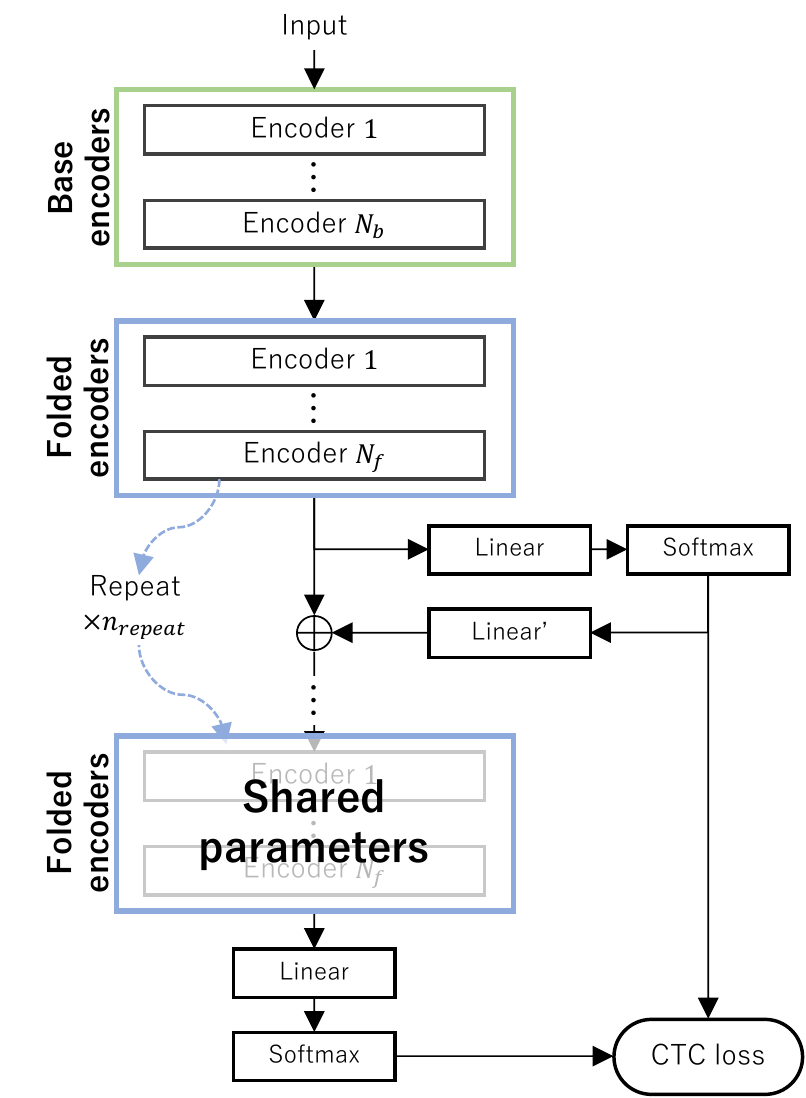}
\caption{
Block diagram of the proposed method.}
\label{fig:block}
\hspace{-5mm}
\end{figure}
\begin{table*}[]
\footnotesize
\centering
\caption{Comparison of the proposed method with 18 conformer encoders. The proposed method shows comparable performance to self-conditioned CTC with fewer parameters, and the performance increases with the number of base encoders $N_b$. The RTF was measured with LibriSpeech's Test Other set.}
\label{tab:my-table}
\begin{tabular}{@{}llcccccccc@{}}
\hline
\multirow{3}{*}{}    &             & \multirow{3}{*}{\#Params} & \multirow{3}{*}{RTF} & \multicolumn{4}{c}{LibriSpeech 100h}                        & \multicolumn{2}{c}{TEDLIUM2}                         \\ \cline{5-10} 
                     &             &                           &                      & \multicolumn{2}{c}{Dev WER}  & \multicolumn{2}{c}{Test WER} & \multirow{2}{*}{Dev WER} & \multirow{2}{*}{Test WER} \\
                     &             &                           &                      & Clean        & Other         & Clean       & Other         &                          &                           \\ \hline
\multicolumn{10}{l}{\emph{Conventional methods, 18 conformer encoders}}                                                                                                                                                                  \\
\multicolumn{2}{l}{CTC}            & 30.5M                     & 0.058                & 8.1          & 22.0          & 9.2          & 24.0          & 9.3                      & 8.8                       \\
\multicolumn{2}{l}{Inter. CTC}     & 30.5M                     & 0.058                & 7.7          & 21.6          & 7.8          & 21.7          & 9.1                      & 8.7                       \\
\multicolumn{2}{l}{Self-cond. CTC} & 30.6M                     & 0.058                & \textbf{7.0} & 20.5          & 7.3          & 20.5          & \textbf{8.7}             & \textbf{8.2}              \\ \hline
\multicolumn{10}{l}{\emph{Proposed method}}                                                                                                                                                                       \\
$N_b=0, N_f=3$         & (Repeat: $6$)   & 6.8M                      & 0.058                & 9.0          & 24.1          & 9.5          & 25.6          & 10.5                     & 10.1                      \\
$N_b=3, N_f=3$          & (Repeat: $5$)   & 11.6M                     & 0.058                & 7.3          & 20.4          & 7.6          & 21.0          & 9.1                      & 8.5                       \\
$N_b=3, N_f=3$          & (Repeat: $6$)   & 11.6M                     & 0.066                & 7.2          & 20.0          & 7.5          & 20.8          & 9.1                      & 8.4                       \\
$N_b=6, N_f=3$         & (Repeat: $4$)   & 16.3M                     & 0.058                & 7.1          & 20.4          & 7.4          & 20.9          & 9.3                      & 8.6                       \\
$N_b=6, N_f=3$          & (Repeat: $6$)   & 16.3M                     & 0.075                & \textbf{7.0} & \textbf{19.7} & \textbf{7.2} & \textbf{20.1} & 9.0                      & 8.3                       \\ \hline
\end{tabular}
\end{table*}

\begin{table}[]\footnotesize
\centering
\caption{WER for LibriSpeech with varying the number of folded encoders $N_f$.
The number of repeats was set to match 18 layers encoders.}
\label{tab:my-table}
\begin{tabular}{lcccc}
\hline
    & \multicolumn{2}{c}{Dev WER} & \multicolumn{2}{c}{Test WER} \\
    & Clean        & Other        & Clean         & Other        \\ \hline
$N_b=3, N_f=1$ (Rep.:15) & 8.4          & 22.7         & 8.7           & 23.4         \\
$N_b=3, N_f=2$ (Rep.: 8) & 7.6          & 20.9         & 8.0           & 21.9         \\
$N_b=3, N_f=3$ (Rep.: 6) & \textbf{7.2}          & \textbf{20.0}         & \textbf{7.5}           & \textbf{20.8}         \\
$N_b=3, N_f=4$ (Rep.: 4) & 7.4          & 20.4         & 7.5           & 21.1         \\ \hline
\end{tabular}
\end{table}
\hspace{-6mm}

\subsection{Relation to prior works}
A network that repeatedly uses encoders with shared parameters has also been proposed by Universal transformer~\cite{dehghani2019universal}. Also, in ASR, \cite{li2019improving} has a similar structure.
However, these methods use only the last output of the iterations to calculate the loss.
Another difference with the proposed method is that there is no base encoder, and the encoder is applied repeatedly to the input from the beginning.
As a result, these may cause performance degradation.


\section{Experiments}
\subsection{Experimental conditions}
Two datasets were used for the experiments, LibriSpeech~\cite{panayotov2015librispeech} and TEDLIUM2~\cite{rousseau2014enhancing}. 
LibriSpeech is a dataset of utterances from English audiobooks, and we used the 100-hour subset for training. 
TEDLIUM2 contains utterances from English Ted Talks, and we used the 210-hour training data. 
We used the standard validation and test sets for model selection and evaluation, respectively. 
The validation and test sets of LibriSpeech are divided into ``clean'' and ``other'' based on the quality of the recorded utterances. 
We used 500 subwords for tokenizing output texts, which were constructed from each training set using SentencePiece~\cite{kudo2018subword}. 
As input speech features, 
we extracted 80 dimentional mel-spectrogram with three-dimensional pitch features using Kaldi~\cite{povey2011kaldi}. 
For data augmentation, we applied speed perturbation~\cite{ko2015audio} and SpecAugment~\cite{park2019specaugment}. 

The Conformer encoder~\cite{gulati2020conformer} is used for the audio encoder layer.
For the self-attention module, 
the number of heads $d_{\mathrm h}$, 
the dimension of a self-attention layer $d_{\mathrm{model}}$, and 
the dimension of a feed-forward network $d_{\mathrm{ff}}$
were set to 4, 256, and 1024, respectively. 
The kernel size of the convolution module was set to 15. 
the encoder consisted of two convolutional neural networks (CNN)-based downsampling layers.

Network parameters were trained with 50 epochs using the Adam optimizer~\cite{kingma2015adam} with $\beta_1\!=\!0.9$, $\beta_2\!=\!0.98$, $\epsilon\!=\!10^{-9}$, and 
the Noam learning rate scheduling~\cite{vaswani2017attention}. 
Warmup steps were set to 25k, and a learning rate factor was 1.0. 
Regularization hyperparameters, such as dropout rate and label-smoothing weight, were the same setup as in~\cite{guo2021recent}.
For evaluation, 
a final model was obtained by averaging model parameters over 10 checkpoints with the best validation performance. 
For decoding, we performed the best path decoding of CTC~\cite{graves2006connectionist}.
All of the decodings were done without using external language models (LMs). 
\begin{figure}[tb]
\centering
\includegraphics[width=0.9\linewidth]{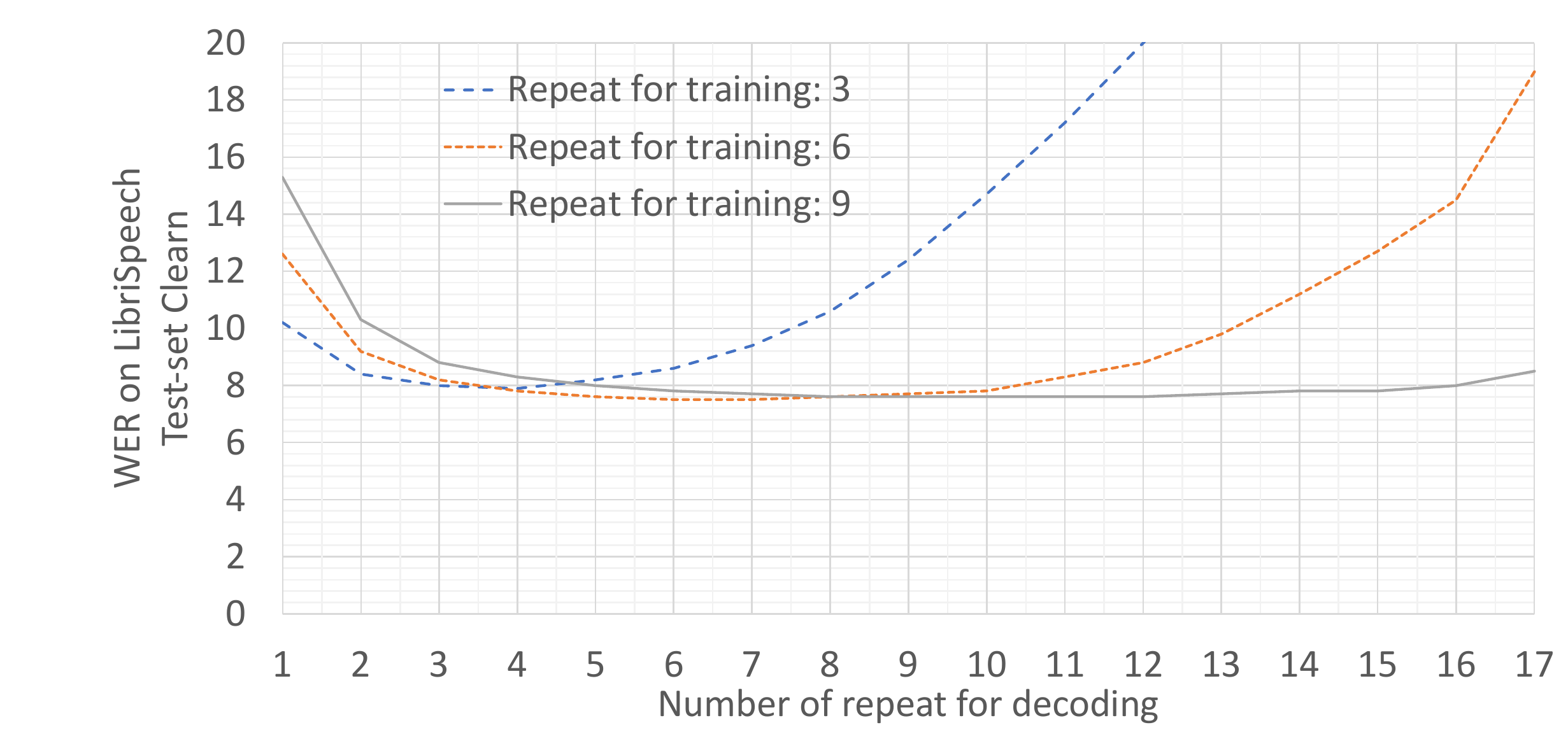}
\caption{Relation of the number for repeats.}
\label{fig:repeat}
\end{figure}
\subsection{Results}
Table~1 shows the comparison results between the proposed and conventional methods when using 18-layer conformers.
It also shows the results of the proposed method when $N_b$ is varied.
Two cases are shown for the number of repeats: one is when the operation is matched to the 18-layer conformer,
and the other is when the number of repeats is fixed at 6.
For the number of base encoders $N_b$, the results for $N_b=0$ were inferior to CTC, but for $N_b=3$ and $6$, the results were comparable to self-conditioned CTC with significantly fewer parameters.
Also, when the number of repeats is set to 6, the CTC performs better than the 18-layer conformer.
This result is reasonable because $N_b=6,N_f=3$ with 6 repeats corresponds 24 layers of operations,
but in terms of training, the proposed method with fewer parameters is effective because 24-layer conformer has more than 40M parameters and requires extensive GPU resources, 

Table 2 shows the WER of librispeech at different $N_f$.
The number of folded encoders is not necessarily large, and the best performance was obtained with 3.
From these results, it is better to obtain `stable' representations with more base encoders before iterating with the folded encoders. 
Once the representation is obtained, it is sufficient to repeat the folded encoder with fewer layers.

\subsection{Number of repeats}
Figure 2 shows the relationship between the number of repeats during training and decoding.
The number of encoder layers were set to $N_b=3$ and $N_f=3$.
It can be seen that when the number of repeats during decoding is larger than the number of repeats during training, the performance gets worse and worse.
It would be desirable to have approximately the same number of repeats during training and decoding.
When the training repeat is as large as 9, it can handle many repeats during decoding, but the best performance is the same as when the training repeat is 6. 
It seems that around 6 for the number of training would be better in this experiment.

\section{Conclusion}
This paper proposed CTC-based non-autoregressive ASR with self-conditioned folded encoders.
We investigated how to set the number of layers and the number of iterations for the base and folded encoders and confirmed that the proposed method achieves a performance comparable to that of the conventional method using only 38\% as many parameters.


\vfill\pagebreak

\ninept
\label{sec:refs}

\bibliographystyle{IEEEbib}
\bibliography{strings}

\end{document}